# JOURNALISTS KNOWLEDGE AND UTILISATION OF GOOGLE TRANSLATE APPLICATION IN SOUTH EAST, NIGERIA

**Onuegbu, Okechukwu Christopher**
Department of Mass Communication, University of Nigeria, Nsukka.
onuegbu.okechukwu.pg89370@unn.edu.ng

**Anorue, Ifeanyi Luke**
Department of Mass Communication,
University of Nigeria, Nsukka.
ifeanyi.anorue@unn.edu.ng

**Oghwie, Bettina Oboakore**
Department of Mass Communication, University of Delta, Agbor.
bettina.oghwie@unidel.edu.ng

**Abstract**

*This study was aimed at finding out if journalists in South East Nigeria have knowledge of Google Translate Application and also utilise it. It adopted a survey design with a sample size of 320 which was determined using Krejcie & Morgan (1970). Its objectives were to ascertain the extent journalists in South East Nigeria know about Google Translate Application, assess the utilisation of Google Translate Application among journalists in South East Nigeria, and identify the challenges affecting the journalists in South East Nigeria while using Google Translate Application. The theoretical underpin was Knowledge Attitude and Practise Model (KAP). The findings showed that journalists in South East Nigeria have knowledge of Google Translate Application but apply it mostly outside the region. It concludes that journalists in South East Nigeria have the knowledge of the App. but apply it outside the zone. The study recommends increased usage of the App. within South East Nigeria.*

**Introduction**

Language is one of the basic elements or components of culture which offers people of the same cultural background the opportunity to interrelate and communicate to one another. It is expressive and receptive, and involves listening, understanding, speaking, using of signs and symbols, gestures and writing. Some experts see it as an identity, identifier and definer of people's culture (Crystal, 2021; Reyes, 2020).

Perhaps, this is why languages are taught in schools and homes. People from various cultural backgrounds learn different languages so they can easily fit in wherever they are or travel to, relate and communicate effectively with others. Some also act as language interpreters, translators or instructors to strangers, visitors and foreigners (Lead, 2021).

However, the Internet, a product of technological advancement powered by Transmission Control Protocol/Internet Protocol (TCP/IP), has made it easier for individuals, at their own comfort, to learn and translate different documents into various

159

languages (Techtarget, 2020). Some of the Internet or online applications purposely designed and developed to perform these tasks are Google Translate App., TripLingo, Document Translator, iTranslate Voice 3, SayHi, TestGrabber, Microsoft Translator, MyLingo, and Wago (Uncubed, 2019).

The Google Translate App, also known as Google Translator application, is the most popular of these online translation machines (Collins, 2023). Developed by Google in 2006, the App as at June 2023, is capable of translating a total of 133 languages with records of over 500 million users and over 100 billion daily translated words (Google Translate, 2023). Among them are Igbo, Hausa, Afrikaans, English, Arabic, Basque and Russian. There are also about 103 languages still being developed by the App (McCamy, 2020).

It works on iOS and Android devices, including iPhone and iPad, but could equally be directly accessed online (on the internet) or offline (after being installed) on mobile phones, laptops and others (Caswell, 2022; Google Translate, 2023; Txabarriaga, 2023). To access it, a user is required to compute already typed texts or type directly on the space provided on the App. It equally translates spoken language into some selected languages or texts contained in a picture (image), as well as assist users with correct pronunciation of words. For websites, the App helps online users to access webpages and data published on them in the users preferred languages (Gestanti, et. al 2019).

Therefore, it is useful to all, especially journalists, who always travel from one place to another interacting with people at different points in time. Journalists owing to the nature of their profession could sometimes travel from English speaking areas to non-English speaking areas. Hence, this research seeks to ascertain if journalists in South East Nigeria have knowledge of Google Translate Application and also utilise it.

Language barrier is considered a problem everywhere in the world because communication is an integral part of society. How can you communicate or express your thoughts, feelings, actions or reasons to anyone at any particular point in time if you do not understand their language? This was probably why Google invented 'Google Translate Application' in 2006 to solve communication problems, among others (Chomsky, 2002). Several scholars have shown how Google Translate App. assisted people in communicating effectively with others who speak different languages other than theirs. So it could not be an understatement to state that the App. has come to rescue people, especially strangers and travellers to foreign lands, places, nations or countries whose languages they do not understand. In Nigeria every state is like a foreign land because the country consists of over 500 languages, more than 300 ethnic groups or nationalities, diverse cultures, different religions and regions (Green, 2023). There are also some Nigerians, including immigrants, finding it impossible to communicate effectively in any other language other than theirs. There are also English speaking-Nigerians who are facing challenges of expressing themselves to non-English speaking people either while travelling or when surfing through the Internet for socialisation, to access important documents, humans or materials (Akinyemi, 2022; Ohai, 2017; Campbell, 2020). Nigerian journalists are in this category of people because their profession always exposes them to both humans and material resources. However, there is little or no information to showcase if journalists in South East Nigeria have knowledge of Google Translate Application and also utilise it. Hence, this study is necessary.

**1.1 Objectives of Study**
The objective of the study was to ascertain if journalists in South East Nigeria are knowledgeable in the use of Google Translate Application and also utilise it in cross-cultural communication. Specifically, this study aims:
1. To ascertain the extent journalists in South East Nigeria know about Google Translate Application;
2. To assess the utilisation of Google Translate Application among journalists in South East Nigeria;
3. To identify the challenges affecting the journalists in South East Nigeria while using Google Translate Application.

**1.2 Research Questions**
The following research questions will be answered in this study:
1. To what extent do journalists in South East Nigeria know about Google Translate Application?
2. How do journalists in South East Nigeria utilise Google Translate Application?
3. What are the challenges affecting the usage of Google Translate Application among journalists in South East Nigeria?

**2.0 Literature Review**
Google Translate App is a revolution in cross cultural communication. With over five hundred million active users and 100 billion daily usage as at June 10, 2023, the technological translation software is dominating cyberspace. It translates all kinds of documents both in local and international languages (Brown, 2021; Caswell, 2022; Google Translate, 2023; More, 2019).

Hence, people viewed it as a blessing while travelling to societies or countries whose languages they do not understand (Moorkens, 2023). Travellers (and other users of the App) download the App on their mobile phones and other technological devices, and use it effectively because it has a record of about 94% accuracy (Collins, 2023). Jacobs (2018), an English-speaking journalist, who enjoyed it during his six months' sojourn to Hong Kong, Singapore, Greece, Israel and Russia, recounts its usefulness thus;

*"While I was there, fans from all over the world were holding up smartphones and tablets to one another to carry out conversations from Russian to English, Spanish to Portuguese, Arabic to French, and every other language pair you can imagine. I used it to translate signs written in the Cyrillic alphabet, talk with taxi drivers, figure out what I had just ordered, and read museum placards… as I walked through a supermarket in*

*Tokyo's Shibuya neighbourhood and encountered one unfamiliar food after another, the camera translated each item before my eyes. It was like putting on glasses for the first time."*

Dahmash (2020), with a Focus Group discussion and interview, found that Google Translate Application could serve as a language learning resource to students, as well help them to improve their vocabularies. Tsai (2019) study also found that Google Translate App. empowers students to improve positively in their English writing by recording fewer errors in the grammar and contents, and vocabulary in Saudi Arabia.

Almufawez and Maroof (2018) in a mixed method study involving a control group and experimental group made up of six female students, found that Google Translate influences student's learning positively especially on comprehension, word selection, and punctuation. Bahri and Mahadi (2016) also found that Google Translate helps students to improve on their vocabularies, writing, and reading.

**2.1 Theoretical Framework**
The theoretical underpin for this study is the Knowledge Attitude and Practise Model (KAP). KAP, propounded around the 1950s, was employed for this study. The theory helps to explain how people feel, behave or respond to new knowledge or act. These include their actions and inactions towards new topics (Andrade et. al, 2020; Kaliyaperumal, 2004). The theory will enable us to understand how Nigerian journalists from the South East region respond, feel, behave, accept or reject Google Translate Application being a new technological innovation. In other words, the theory will tell if the journalists are aware (have knowledge of Google Translate App), accept or expose (take actions) and apply or utilise (practise) what they have learnt from the App.

**3.0 Methodology**
**Research Design**
The research design adopted in this study is a sample survey: a descriptive study and an investigation in which only part or sample of the population is studied and selection is made in a true representation of the whole population.

**Area of Study**
The area of study is the South East geographical zone or region of Nigeria. The zone comprises five States, namely Abia, Anambra, Ebonyi, Enugu and Imo. Igbo is the major ethnic nationality and language in the region although there are several other ethnic nationalities and languages spoken by the residents such as English, Igala, Hausa, Yoruba and Pidgin-English. Also, there are many radio stations, newspapers companies, television stations, the new media, including practising journalists and other media practitioners in the zone (Onuegbu, 2021). In the constitution of the Nigeria Union of Journalists (NUJ), States in the South East are grouped under Zone C with five Councils (chapters) and 74 Chapels (units/subgroups). Abia State Council has 5 chapels, Anambra State Council (21 chapels), Ebonyi State Council (16 chapels), Enugu State Council (16 chapels), and Imo State Council (16 chapels).

**Study Population**
The population of the study are all practising journalists in South East Nigeria: males and females. Official data obtained from the NUJ Zone C respective States Councils showed that there are 1,863 registered practising journalists in South East, Nigeria (Ifesinachi, 2013).

**Sample Size**
The sample size for the study was 320 which was determined using Krejcie & Morgan (1970) sample table. This was chosen because 1,863 registered practising journalists are a finite population.

**Sampling Procedure**
The sampling procedure was multi-stage in approach. The researcher conducted his sampling in stages. At the first stage, journalists in South East Nigeria were divided in clusters using the existing structures of the NUJ (State Councils). Their meeting areas or Councils headquarters were identified as Umuahia, Awka, Abakaliki, Enugu and Owerri. The meeting days otherwise referred to as Congress days of the respective Councils were further established as once monthly. The researchers employed the services of three research assistants to assist in administering questionnaires to journalists in Imo, Ebonyi and Abia States Councils, while two of them did the same thing at Anambra and Enugu State councils respectively. 64 respondents were randomly selected from each of the congresses, making it a total of 320.

**Instrument of Data Collection**

Questionnaire was used to generate data for this study. The questionnaire was closed-ended.

**Validity and Reliability Test of instrument**

Face-to-face validity was used to validate the questionnaire used in this study. Two lecturers from the Department of Mass Communication at the Nnamdi Azikiwe University, Awka and University of Delta, Agbor, as well as an Editor in one of the weekly newspapers in Awka, Anambra state, were consulted for both face and content validity. Their advice was incorporated into the research instrument. To assess reliability, a test-retest method was employed, where the instrument was administered to 20 respondents (ten from each university) and then re-administered after ten days. The obtained results were subjected to a reliability test using Pearson's Correlation Coefficient formula proposed by Francis Galton in 1880. The analysis convinced the researchers that the instrument was reliable because there was no significant difference between the answers supplied by the respondents at both the first and the second.

**Method of Data Analysis**
The researcher's method of data analysis was quantitative. Answers extracted via the questionnaire were recorded as numeral data. The frequency of each answer was found and the percentage computed accordingly using charts, before the researcher proceeded to interpret them towards answering the research questions.

**4.0 Data Presentation and Analysis**
Out of the 320 questionnaire copies distributed, 279 representing 87% were recovered while 41 representing 13% were not. Hence, the response rate stood at 279 (87%).

**Demographic Data**
The demographic data showed that 163 of the respondents representing 58% were males and 116 respondents (42%) were females. Also, the respondents under the ages of 18-28 were 77 (28%), 29-39 were 90 (33%), 40-50 were 71 (25%), and 51 and above were 41 (14%). The respondents marital status indicates that 122 or 44% were single, 125 or 45% were married, 9 (3%) were divorcees, and the widows/widowers were 23 or 8%. Similarly, the educational background of the respondents showed that those who possessed either a Degree (B.Sc./B.A/B.Tech.) or Higher National Diploma (HND) were 109 (39%), Ordinary National Diploma (oND) and National Certificate Education (NCE) were 80 (29%), Masters' (MSc./M.A) and Postgraduate Diploma were 77 (28%), and PhD 13 (5%). The professional destinations of the respondents were Reporters (85 or 30%), Correspondents (111 or 40%), Sub-Desk (33 or 12%) and Editors (50 or 18%). Again, the respondents' media were categorised under Print-102 or 37%, Electronic-89 or 32% and Online/New Media 88 or 31% respondents.

**Table 1: Knowledge of Google Translate Application**

| Have you heard about Google Translate Application | Frequency | Percentage |
|---|---|---|
| Yes | 238 | 85% |
| No | 41 | 15% |
| **TOTAL** | 279 | 100 |

The data in the table one above shows that 238 respondents representing 85% have heard about Google Translate Application, while 41 (15%) respondents feigned ignorance of the Application.

**Table 2: Knowledge level of Google Translate App.**

| Functions of Google Translate Application | Frequency | percentage |
|---|---|---|
| Voice translation | 46 | 20% |
| Picture translation | 45 | 19% |
| Video translation | 41 | 17% |
| Text translation | 39 | 16% |
| Website | 27 | 11% |
| All of the above | 40 | 17% |
| Total | 238 | 100% |

In Table 2, the functions of Google Translate App. are voice translation (46 or 20% of the respondents), picture translations (45 representing 19% of the respondents), video translations-41 respondents representing 17% and all of the above-40 respondents representing 17%. Others were text translations (39 respondents representing 16%) and website translations (27 respondents representing 11%).

**Table 3: Utilisation of Google Translate Application**

| Have used Google Translate Application | Frequency | Percentage |
|---|---|---|
| Yes | 74 | 31% |
| No | 117 | 49% |
| Not Sure | 47 | 20% |
| Total | 238 | 100% |

The data in Table 3 shows that 74 respondents representing 31% have utilised Google Translate Application, 117 or 49% have not used it, while 47 or 20% of the respondents are not sure if they have used it or not.

**Table 4: Level of use of Google Translate App.**

| How often respondents used Google Translate App. | Frequency | Percentage |
|---|---|---|
| Always | 24 | 32% |
| Very often | 0 | 0% |
| Sometimes | 27 | 37% |
| Rarely | 23 | 31% |
| Total | 74 | 100% |

Table 4 shows that of 74 respondents who proved usage of Google Translate Application, 27 representing 37% have used the Google Translate Application sometimes, always (24 representing 32%), rarely (23 representing 31%) and very often- 0 (0%).

**Table 5: Extent of use of Google Translate Application**

| Reasons for using Google Translate App. | Frequency | Percentage |
|---|---|---|
| Research or interactions on the Internet | 37 | 50% |
| Offline interactions with people or objects within South East Nigeria | 4 | 5% |
| Offline interactions with people or objects while travelling to other parts of Nigeria | 13 | 18% |
| Interactions with people while travelling outside Nigeria | 20 | 27% |
| Total | 74 | 100 |

Data in Table 5 shows that 37 respondents (50%) have used Google Translate application for research or interactions on the Internet, 20 (27%) of the respondents have done that while travelling and interacting with people outside Nigeria, 13 (18%) of the respondents did it while travelling and interacting with people and objects in other parts of Nigeria, while only 4 (5%) of the respondents have used it for offline interactions or objects within South East Nigeria.

**Table 6: Challenges facing use of Google Translate Application**

| Factors militating against use of Google Translate App. | Frequency | Percentage |
|---|---|---|
| Power failure | 27 | 36% |
| Poor Internet connectivity | 15 | 20% |
| Insufficient knowledge of how it works | 14 | 19% |
| Expensive data/Airtime | 6 | 9% |
| All of the above | 14 | 19% |
| Total | 74 | 100% |

Table 6 reveals the challenges affecting the usage of Google Translate Application as stated by 74 respondents. The challenges are power failure (27 or 36%), poor Internet connectivity (15 or 20%), All of the above (14 or 19%), insufficient knowledge of how it works (12 or 16%), and expensive airtime/data (6 or 9%).

**4.1 Discussion of finding**
The study investigated the knowledge and utilisation of Google Translate Application among Journalists in South East, Nigeria. It was ascertained that journalists practising in South East Nigeria have knowledge of Google Translate Application (App.). Result on Table 1 shows that up to 85% of the respondents are aware of Google Translate App. The study also found out that the Google Translate Application has been used for voice translation, picture translations, video translations, text translations and website translations. This was not different from the conclusion of McCamy (2020) who explained that Google translate App is popular. However, the assessment showed that journalists in South East Nigeria are not adequately utilising the Google Translate Application as 49% of the respondents (in Table 3) said they have not done so as against 31% that have utilised it. Also, data in table four showed that only 32% of the respondents are utilising the Google Translate App. always. It was also found that most (50%) of journalists in South East Nigeria are using Google Translate Application for research or interactions on the Internet. The study identified the challenges affecting or contributing to poor usage of Google Translate Application among journalists in South East Nigeria to include power failure (36%), poor Internet connectivity (20%), insufficient knowledge of how it works (16%), and expensive airtime/data (9%). These findings were not different from the findings of Bahri and Mahadi (2016), Dahmash (2020), and Vieira, et. al (2020).

**5.0 Conclusion**
From the results of the study, we conclude that journalists in South East Nigeria have knowledge of Google Translate Application but are using it mostly on the Internet. It was therefore concluded that Google Translate App. is underutilised among journalists in South East Nigeria.

**5.1 Recommendations**
Following the above findings, it is recommended that:

(i) There is a need for media managers, Nigeria Union of Journalists (NUJ) or nongovernmental organisations to organise a mass awareness campaign and training to intimate and build capacities of journalists in South East Nigeria on uses and effectiveness of Google Translate App.
(ii) The journalists in South East Nigeria are also encouraged to be using Google Translate App. within the zone as they use it on Internet and other places.

(iii) Developers of Google Translate App. should keep on updating the applications with relevant modern features, facilities and languages to make it fit in interpreting different languages, signs and symbols appropriately.